\documentclass[twocolumn,prc,aps,showpacs,superscriptaddress,nofootinbib]{revtex4-1}
\usepackage{amsmath}
\usepackage{amssymb}
\usepackage{graphicx}
\usepackage[usenames]{color} 
\usepackage{xcolor}
\usepackage{ulem}

\begin{document}
\title{Correlations between charge radii differences of mirror nuclei
	and stellar observables}
\author{P. Bano}
\affiliation{School of Physics, Sambalpur University, Jyotivihar-768 019, India.}
\author{S. P. Pattnaik}
\affiliation{School of Physics, Sambalpur University, Jyotivihar-768 019, India.}
\author{M. Centelles}
\affiliation{Departament de F\'isica Qu\`antica i Astrof\'isica (FQA),
	Universitat de Barcelona (UB), Mart\'i i Franqu\`es 1, E-08028 Barcelona, Spain}
\affiliation{Institut de Ci\`encies del Cosmos (ICCUB),
	Universitat de Barcelona (UB), Mart\'i i Franqu\`es 1, E-08028 Barcelona, Spain}
\author{X. Vi\~nas}
\affiliation{Departament de F\'isica Qu\`antica i Astrof\'isica (FQA),
Universitat de Barcelona (UB), Mart\'i i Franqu\`es 1, E-08028 Barcelona, Spain}
\affiliation{Institut de Ci\`encies del Cosmos (ICCUB),
Universitat de Barcelona (UB), Mart\'i i Franqu\`es 1, E-08028 Barcelona, Spain}
\affiliation{Institut Menorqu\'i d'Estudis, Cam\'i des Castell 28, 07702 Ma\'o, Spain}
\author{T. R. Routray}
\affiliation{School of Physics, Sambalpur University, Jyotivihar-768 019, India.}
\date{\today}
\begin{abstract}
 The correlation between the charge radii differences in mirror nuclei pairs and the neutron skin thickness 
 has been studied with the so-called finite range simple effective interaction over a wide mass region.
 The so far precisely measured charge radii difference data within
their experimental uncertainty ranges in the
$^{34}$Ar-$^{34}$S, $^{36}$Ca-$^{36}$S, $^{38}$Ca-$^{38}$Ar, and $^{54}$Ni-$^{54}$Fe mirror pairs are used to
ascertain an upper limit for the slope parameter of the nuclear symmetry energy L$\approx$ 100 MeV. 
 This limiting value of L is found to be consistent with the upper bound of the NICER
PSR J0740+6620 constraint at 1$\sigma$ level for the radius R$_{1.4}$ of 1.4M$_\odot$ neutron stars.
The lower bound of the NICER R$_{1.4}$ data constrains the lower limit of L to $\approx$70 MeV. Within the range for
L=70--100 MeV the tidal deformability $\Lambda^{1.4}$ constraint,  which is extracted from the GW170817 event
at 2$\sigma$ level, and 
the recent PREX-2 and CREX data on the neutron skin thickness are discussed.
\end{abstract}

\maketitle

\section{Introduction}
With the advent of collinear laser beam technology \cite{Yang2023,Campbell2016}, the precise measurement of nuclear charge radii in the magic nuclei 
isotopic chains for the medium-heavy mass region ranging between Z=19 to Z=50 could be completed with the recent measurements of the Ni-isotopic 
chain (\cite{Malbrunot-Ettenauer2022} and references therein). Due to this high-precision technology, the charge radii difference, $\Delta$R$_{CH}$, 
between the two nuclei of a mirror pair
can be measured accurately. The experimental determination of $\Delta$R$_{CH}$ of mirror pairs provides the opportunity of estimating the neutron skin thickness in nuclei, 
$\Delta$R$_{np}$, under the isospin symmetry of the nucleon-nucleon (NN) interaction \cite{Brown2017,Yang2018,Sammarruca2018,
	Brown2020,Gaidarov2020,Pineda2021,ReinhardL2022,Novario2023}. Taking into account the charge symmetry breaking effects,
	 $\Delta$R$_{CH}$ of mirror nuclei appears to be a clean (electromagnetic) 
alternative method to the electroweak parity-violating asymmetry \cite{Abrahamyan2012} and electric dipole polarizability
\cite{Tamii2011,Piekarewicz2012,Rossi2013} measurements for the estimation of the neutron skin thickness.
The neutron skin $\Delta$R$_{np}$ in a nucleus is defined as 
the difference between the root-mean-square ({\it {rms}}) neutron and proton radii (R$_n$-R$_p$). 
 Given that $\Delta$R$_{np}$ of neutron-rich nuclei is strongly correlated to the slope of the symmetry
energy $L$ \cite{Brown2000,Reinhard2010,Thiel2019,Piekarewicz2019}, the knowledge of $\Delta$R$_{np}$ is very valuable for
constraining the equation of state (EoS) of isospin-asymmetric matter. 
However, neutron radii of nuclei, and hence neutron skins, are more difficult
to determine with high accuracy than charge radii.
In mirror nuclei the measurement of the neutron radius may be bypassed if one takes into account that
under perfect charge symmetry of the NN-interaction, the neutron radius
of the nucleus with Z protons and N neutrons equals the proton radius
of its mirror pair with N protons and Z neutrons.
Then, the neutron skin of a (Z,N) nucleus can be obtained from the charge radii of the mirror pair 
as $\Delta$R$_{np} = \Delta$R$_{CH} \equiv$ R$_{CH}$(N,Z)$-$R$_{CH}$(Z,N) \cite{Brown2020,Pineda2021}.
In reality, the Coulomb interaction breaks the perfect charge symmetry and the 
equality between $\Delta$R$_{CH}$ and $\Delta$R$_{np}$ is weakened. 
It is therefore important to study the degree of correlation preserved between $\Delta$R$_{CH}$ and $\Delta$R$_{np}$,
 and between $\Delta$R$_{CH}$ and the symmetry energy slope~$L$.

An interesting current trend in the literature is to use the neutron skin property of finite nuclei to extract information on the L-value, 
i.e., the pressure in pure neutron matter, and use this to study astrophysical observables under certain assumptions 
\cite{Fattoyev2018,Brown2020,Reed2021}. The measurement of the neutron skin thickness in finite nuclei has important implications for neutron star (NS) properties such as the radius and the tidal deformability of NSs
\cite {Fattoyev2018} and complements the gravitational wave (GW) observations \cite{Zhang2020,Guven2020,Baiotti2019,Piekarewicz2019,Tsang2019PLB,Fasano2019}. Also, recent studies have pointed out
the possible correlation between the mirror pair charge radii difference $\Delta$R$_{CH}$ and the radius of NSs \cite{Yang2018}.
But invariably the range for L extracted from the 
correlation of the terrestrial data and celestial observables could barely explain 
the PREX-2 result of $\Delta R{_{np}^{208}}$=0.283$\pm$0.071 fm for the neutron skin thickness in $^{208}$Pb,
obtained in a model-independent way from parity violating electron scattering (PVES) measurements \cite{Adhikari2021}. 
Moreover, the recently published CREX result of $\Delta R{_{np}^{48}}$=0.121$\pm$0.026(exp)$\pm0.024$(model) fm in $^{48}$Ca \cite{Adhikari2022} from a similar PVES experiment 
suggests the question of why the skin thickness in $^{48}$Ca is so thin. 
Although this low value of the skin in $^{48}$Ca has been predicted by  chiral effective 
field theory (EFT) calculations \cite{Hagen2016,simonis2019}, the PREX-2 and CREX measurements together pose a challenge to the theoretical models for the simultaneous reproduction of both data under 
the present scenario of incomplete
knowledge of the EoS of neutron-rich nuclear matter, as discussed in 
Refs.\cite{Reinhard2022,Paar2023,Tagami2022} within the  Energy Density Functional (EDF) approach.
The computation of $^{208}$Pb from {\it{ab initio}} theory
has not been feasible for now, but combined with advances in quantum many-body techniques has predicted the neutron skin 
in $^{208}$Pb to be $\Delta R{_{np}^{208}}$=0.17$\pm$0.05 fm at 90\% confidence level \cite{Hu2022},
and a value in $^{48}$Ca of $\Delta R{_{np}^{48}}$=0.16$\pm$0.04 fm at the same confidence level \cite{Hu2022}.

In this work, using the so-called finite range simple effective interaction (SEI) model we examine the global correlation
between $\Delta$R$_{CH}$ and $\Delta$R$_{np}$ and show a linearity between them, confirming the trends
previously found in Skyrme forces \cite{Brown2017,Gaidarov2020} and in chiral EFT calculations in the low-energy limit \cite{Sammarruca2018}. In former studies we 
have shown that SEI accurately satisfies a wide range of
constraints in nuclear matter and nuclear structure \cite{Behera1998,Behera2013,Behera2015,Routray2021,Routray2022}.
Also SEI has been used in the astrophysical domain to derive the mass-radius 
relation in NSs as well as dynamical properties, such as the r-mode oscillation and spin-down features associated
with NS physics \cite{trr2007,trr2016,trr2018,mario2019,trr2021}.
 Here, we next use the SEI model to study the experimental $\Delta$R$_{CH}$ data available for four mirror pairs. 
We find that these accurate experimental results, together with the NICER 
telescope data, which constrains the radius of 1.4M$_\odot$ NSs from observations on pulsar PSR J0740+6620 \cite{Miller2021}, predict from the analysis with SEI that the symmetry energy slope L lies in a range between 70 and 100 MeV.
The validity of the L range thus obtained is examined in the context of the tidal 
deformability constraint 
extracted from the GW170817 detection of gravitational waves from a binary NS merger \cite{Abbott2018}, together with the results from the PREX-2 \cite{Adhikari2021} and CREX \cite{Adhikari2022} experiments.

\section{Basic Theory}
The SEI was introduced in Ref. \cite{Behera1998} by Behera and collaborators. 
Its explicit form, for a Gaussian finite range form factor, in coordinate space is given by  
%
\begin{eqnarray}
\lefteqn{V_\textrm{eff} = }\nonumber \\
& & t_{0}(1+x_{0}P_{\sigma})\delta(\vec{r}) +\frac{t_{3}}{6}(1+x_{3}P_{\sigma})\left(\frac{\rho(\vec{R})}
{1+b\rho(\vec{R})}\right)^{\gamma}\delta(\vec{r})\nonumber \\
& & \mbox{}+ (W+BP_{\sigma}-HP_{\tau}-MP_{\sigma}P_{\tau})e^{-r^{2}/\alpha^{2}}\nonumber\\
& & \mbox{} + \text {Spin-orbit part},
\label{SEI}
\end{eqnarray}
where a zero-range spin-orbit (SO) interaction depending on a strength parameter $W_0$ is taken to deal with finite nuclei. All the 
parameters of the interaction, except $t_{0}$ and the SO strength $W_0$, namely, $\alpha, \gamma, b, x_{0}, x_{3},  
t_{3}, W, B, H$, and $M$ are fitted in nuclear matter (NM) of different types using very generic considerations obtained from the 
experimental/empirical conditions \cite{Behera1998,Behera2013,Behera2015}. For example, the density dependence of the isovector part of the EoS is 
fixed from the condition that the asymmetric contribution to the nucleonic part of the energy density in charge neutral beta-stable 
n+p+e+$\mu$ NS-matter is the maximal one \cite{trr2007}. This condition determines the characteristic symmetry energy slope 
parameter L$_C$ for the interaction set and predicts the density dependence of the symmetry energy which is neither very stiff nor very soft 
(see in this respect Figs. 3 and 4a of Ref. \cite{trr2007}). The so-determined L$_C$ is in the range $\approx$75-77 MeV for the 
EoSs of the Gaussian form of SEI in Eq.(\ref{SEI}) corresponding to different $\gamma$ values (different incompressibility), which is in the upper 
ranges for the L-values predicted from the 
theoretical studies of dipole polarizability \cite{Piekarewicz2012,Roca2015}, skin thickness in Sn \cite{Klimkiewicz2007}, {\it {ab
		initio}} calculation of low-density neutron EoS and maximum mass of 2.1M$_\odot$ for NSs \cite{Drischler2020,Essick2021,Tsang2019}.
But the recent studies such as 
charge exchange elastic scattering \cite{Danielewicz2017}, charge radii difference in mirror nuclei \cite{Brown2020}, pion ratio in heavy-ion 
collisions \cite{Estee2021}, and the PREX-2 result on the $^{208}$Pb skin \cite{Adhikari2021}, predict larger values for the L parameter, which are 
safely covered by the L$_C$-values of SEI sets. Also the range for the symmetry energy $E_s(\rho_0)$ at saturation density $\rho_0$ inferred from 
these latter studies is larger than the ones conventionally used in the Skyrme and Gogny interactions. The SEI sets predict $E_s(\rho_0)$
in the range 
$\approx$35-36 MeV, which is in good agreement with the aforementioned experimental results and with the typical values in covariant EDFs.

 The $\rho_0$ and $E_s(\rho_0)$ values in the mentioned range for SEI are the consequence of the parameter fitting protocol
to predict the charge radii of 86 even-even nuclei and the binding energies (BEs) of 161 even-even nuclei with minimum {\it rms} deviations, 
which are obtained in the range of $\approx$0.015 fm and $\approx$1.5 MeV, respectively \cite{Behera2013,Behera2015}.
For comparing with the experimental values, the charge radii were calculated from the relation
R$_{CH}$=$\sqrt{R_{p}^2+0.64 \mbox{fm}^2}$, where $R_p$ is the point proton radius and the factor 0.64 accounts
for the finite-size correction to the point proton distribution \cite{Perera2021}.
Here, one neglects the small additional contributions to the charge radius that arise from the finite size of the neutron
and from relativistic effects \cite{ong2010}. 
In SEI, for a given stiffness $\gamma$ of NM, there is an optimal value of the saturation density $\rho_0$
for which the charge radii of the 86 even-even nuclei used in the fit exhibit
the minimum {\it rms} deviation from the data \cite{Behera2015}.
Therefore, in order to have consistency with the parameters of SEI,
we have computed R$_{CH}$ for the mirror nuclei discussed in this work from the same relation given above.

The full determination of the SEI parameters for symmetric 
and asymmetric NM and for finite nuclei is discussed in Refs.\cite{Behera2013,Behera2015,Routray2021,Routray2022}. 
The SEI is able to reasonably reproduce the microscopic trends of the EoS and the momentum dependence of the mean field 
in NM 
\cite{Behera09,Sammarruca2010,Wiringa1988,APR1998}. 
The finite nuclei are described by using the Quasi-local Density Functional Theory (QLDFT) \cite{Soubbotin2000,Soubbotin2003}. 
To deal with pairing correlations in open-shell nuclei, we use an improved BCS approach, which takes partially into account the 
continuum through quasi-bound states by the centrifugal and Coulomb barriers and allows one to perform pairing calculations near the drip lines 
\cite{Estal2001}. In the SEI model we use a zero-range density-dependent pairing interaction \cite{Bertsch1991}
fitted to reproduce the pairing gaps in NM predicted by the Gogny interaction \cite{Behera2013}.
An excellent agreement between the 
predictions for spherical nuclei using  the QLDFT and the full HF or HFB results including the exchange terms at mean-field 
level was shown
in Ref.~\cite{Behera2016}. 
 It is also found that single-particle energies calculated with SEI are in
reasonable agreement with the experiment and describe the spectrum of $^{208}$Pb with
comparable or better quality than other effective interactions \cite{Behera2013}.
The SEI mean field also reproduces the experimental kink in the isotopic shifts of
the Pb charge radii \cite{Behera2013}, which is not predicted by
other nonrelativistic models, such as Skyrme and Gogny, with an 
isospin-independent SO interaction.
The recent study on the 1$f_{5/2}$-2$p_{3/2}$ proton level crossing and spin inversion phenomena in the Ni and Cu isotopic chains, respectively, using SEI
\cite{Routray2021,Routray2022} 
  has shown that the SEI EoS corresponding to $\gamma$=0.42 is able to produce the experimentally
observed crossing at the right mass number. In this work we use this SEI EoS, whose parameters are given in Table 1 of Ref.\cite{Routray2022} 
and whose  NM saturation properties are reported here in Table {\ref{tab-inm}}
for ready reference.
 It is pertinent 
to note that under the fitting protocol of the SEI parameters, we can relax the condition that determines the characteristic slope parameter value L$_C$ and alternatively impose an arbitrary L-value. The SEI interaction is flexible enough so that this 
change in the fitting procedure does not affect the isoscalar predictions. In the isovector sector, also the n-p effective mass splitting remains invariant. As a consequence, the BEs and {\it {rms}} mass radii of finite nuclei are practically unaffected by the changes of L.
%
\begin{table}[t!]\small
	\centering
	\begin{tabular}{cccccc}
		\hline\hline \rule{0mm}{4mm}
		$\rho_0$    & $e(\rho_0)$ & $K$ & $m^*/m$   & $E_s(\rho_0)$ & $L$ \\
		~(fm$^{-3}$)~ &  ~(MeV)~  &  ~(MeV)~ &        & ~(MeV)~ &~(MeV) \\[1mm]  \hline \rule{0mm}{4mm}
		0.1584     & $-$16  & 240   &  0.711  & 35.5   & 76.71 \\[1mm]\hline\hline
	\end{tabular}
	\caption{Nuclear matter saturation properties of the SEI EoS: density $\rho_0$,
		energy per nucleon $e(\rho_0)$, incompressibility $K$, effective mass $m^*/m$,
		symmetry energy $E_s(\rho_0)$, and slope $L$ of the symmetry energy.}
	\label{tab-inm}
\end{table}

\section{Results and Discussions}
\begin{figure}[t]
	\begin{center}
		\includegraphics[height=9.cm,width=9.5cm]{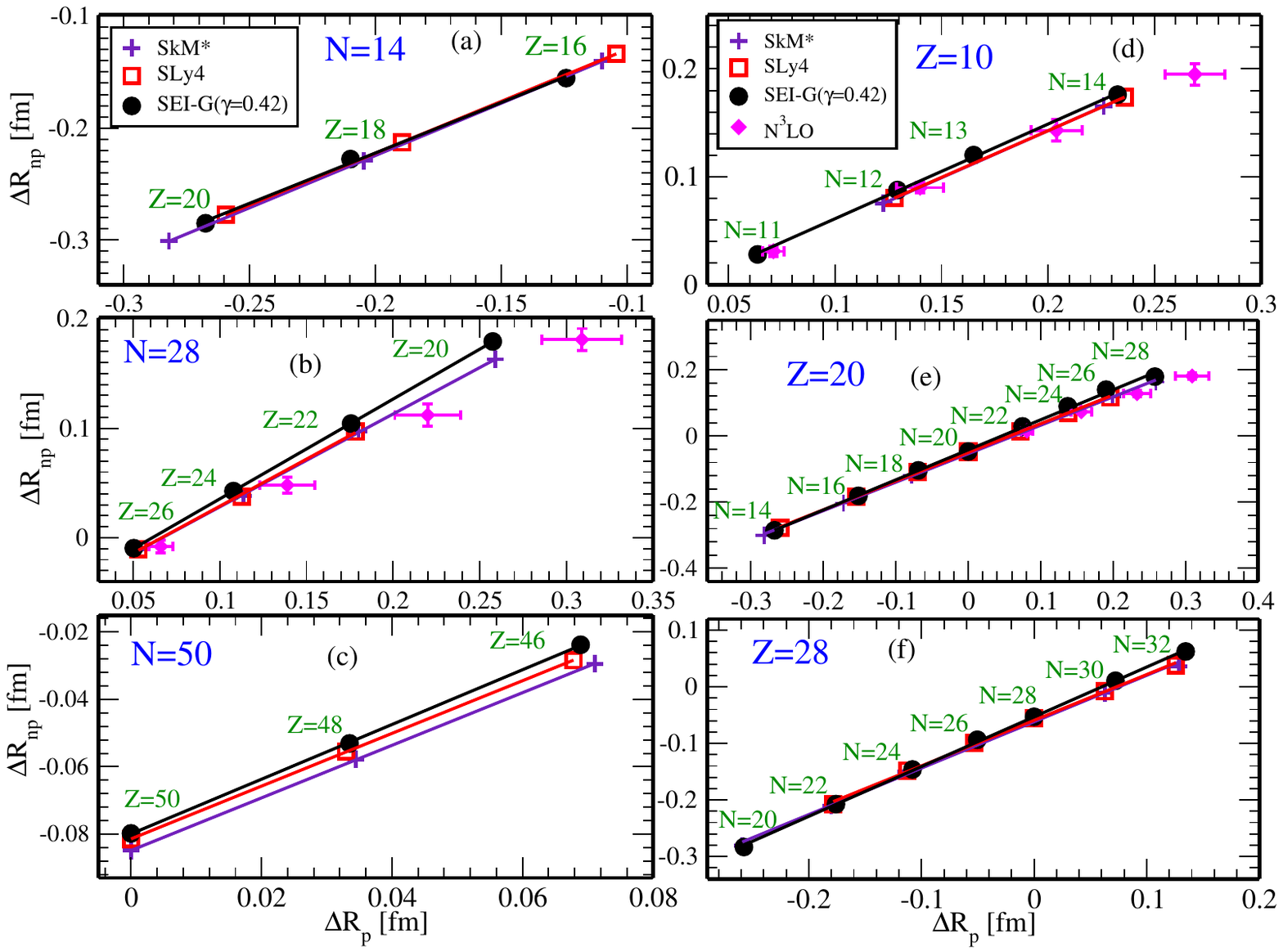}
		\caption{Linear correlation between the neutron skin $\Delta R_{np}$ and the proton {\it {rms}} radii difference $\Delta R_{p}$ 
			for mirror nuclei pairs shown for the SEI EoS in the isotonic chains of (a) N=14, (b) N=28, and (c) N=50 and isotopic chains of 
			(d) Z=10, (e) Z=28 and (f) Z=50. The results of SkM*, SLy4 \cite{HFB_masstable}
			and N$^{3}$LO \cite{Sammarruca2018} 
			forces are given by symbols plus (indigo), square (red) and diamond (magenta), respectively.}
		\label{N14_N28_N50_Z10_Z20_Z28_g42SEIG}
	\end{center}
\end{figure}
The correlation between the skin thickness $\Delta$R$_{np}$ and proton rms radii difference in the mirror pair, 
$\Delta$R$_{p}$=$[R_{p}(N, Z) - R_{p}(Z, N )]$, is examined by computing them in isotonic chains of N=14, 28 and 50, and isotopic chains of Z=10, 
20 and 28, which covers nuclei from light to medium-heavy mass region. 
The results are shown in the six panels of Fig.\ref{N14_N28_N50_Z10_Z20_Z28_g42SEIG}. In each panel the 
results obtained with the Skyrme SLy4 and SkM* \cite{HFB_masstable} sets are also given. In panels (b), (d) and (e) for N=28, Z=10 and Z=20, the chiral EFT results from Ref.\cite{Sammarruca2018} are also shown. 
Using the SEI data given in the six panels we perform a linear fit 
\begin{equation}
\Delta R_{np} = a\,\Delta R_{p}  + b,
\label{Eq_R_np}
\end{equation}
where the values of the coefficients $a$ and $b$ are
\begin{equation}
a=0.881 \pm 0.036 \quad b=-0.049 \pm 0.017 \;{\rm fm}.
\label{Eq_ab}
\end{equation}
Our calculation includes Coulomb and pairing correlations. For similar values of $\Delta R_{p}$, the values of $\Delta R_{np}$ 
are nearly the same. If we measure the proton rms radii difference in a mirror nuclei pair, then from Eqs.(\ref{Eq_R_np}) and (\ref{Eq_ab}) the 
neutron skin of the neutron-rich partner nucleus can be at once estimated. With appropriate consideration of the Coulomb effects as well as other 
symmetry breaking effects, the charge radii difference in mirror pairs can be a surrogate to the weak interaction PVES
experiment for skin thickness measurement. Our present study verifies the $\Delta$R$_{p}$-$\Delta R_{np}$ correlations reported 
previously in Refs.\cite{Sammarruca2018,Gaidarov2020}, which were obtained in the context of chiral EFT and Skyrme interactions. 

 Regarding the experimental information,
data on the charge radii difference, $\Delta R_{CH}$, in mirror pairs 
are available only for a few cases, within maximum neutron and proton number differences of N-Z=4. 
%
\begin{table}[t]\small
	\centering 
		\begin{tabular}{lcccc}
			\hline\hline \rule{0mm}{4mm}
			& \multicolumn{4}{c}{$\Delta R_{CH}$(fm)} \\[0.5mm]
			\cline{2-5} \rule{0mm}{5mm}
			& $^{34}$Ar-$^{34}$S & $^{36}$Ca-$^{36}$S & $^{38}$Ca-$^{38}$Ar & $^{54}$Ni-$^{54}$Fe \\[1mm]
			\hline \rule{0mm}{4mm}
			Expt:~~ &  0.082(9)\:\cite{Brown2017} &  ~0.150(4)\:\cite{Brown2020} & ~0.063(3)\:\cite{Brown2020} & ~0.049(4)\:\cite{Pineda2021} \\[1mm]
			\hline \rule{0mm}{4mm}
			SEI:~ &  0.087~~~~& 0.147~~ & 0.066~~~ & 0.049~~~ \\[1mm] \hline
 \hline
		\end{tabular}
	\caption{Experimental results for the charge radii difference of mirror pair nuclei and the
		predictions of the characteristic SEI EoS having $L=76.71$ MeV.}
	\label{tab-drch}
\end{table}
\begin{figure*}[t]
	\begin{center}
		\includegraphics[width=0.80\textwidth]{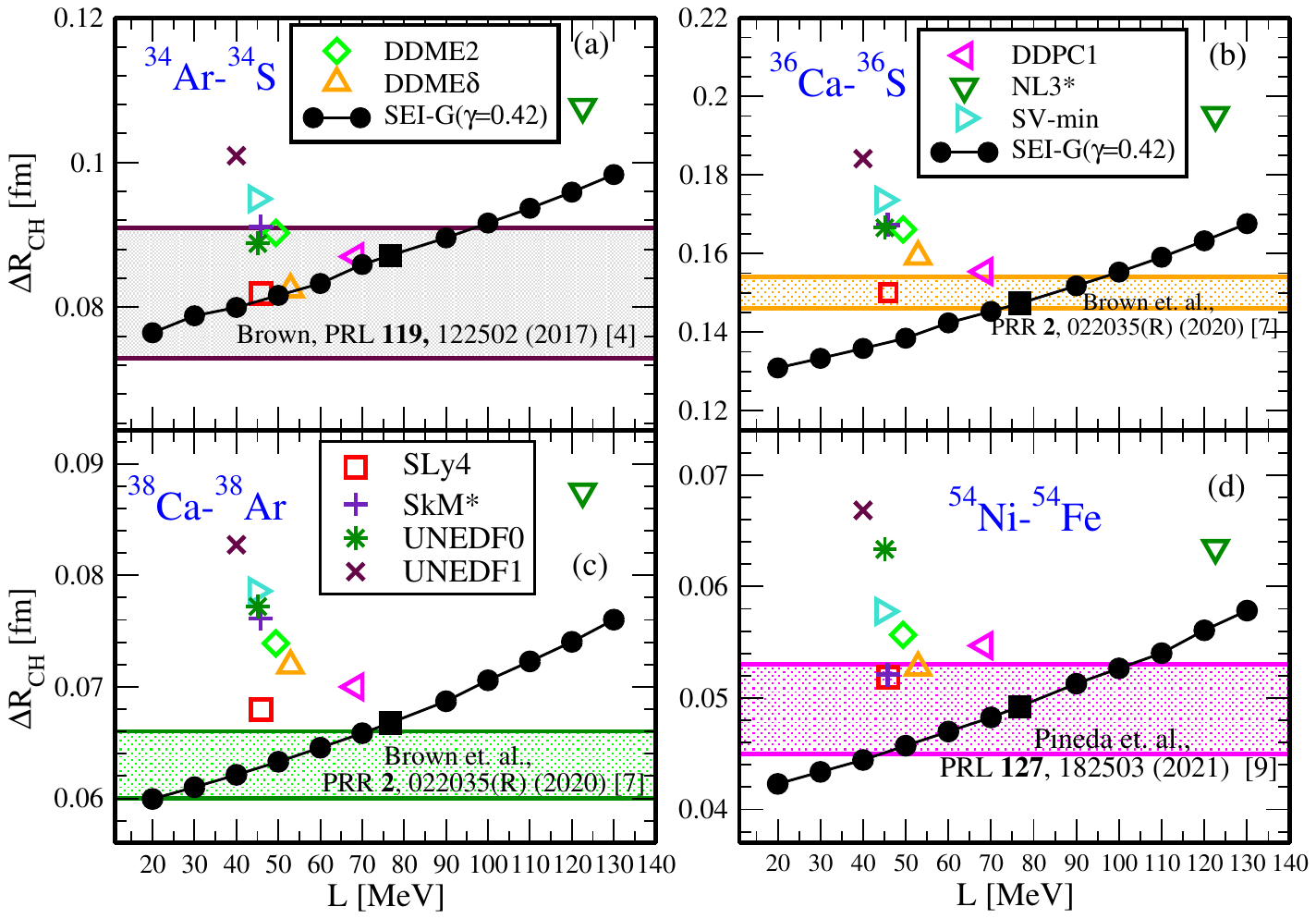}
		\caption{$\Delta R_{CH}$ as a function of the symmetry energy slope L for the mirror pairs (a)$^{34}$Ar-$^{34}$S, (b)$^{36}$Ca-$^{36}$S, (c)$^{38}$Ca-$^{38}$Ar, and (d)$^{54}$Ni-$^{54}$Fe. The experimental results are shown in horizontal bands in each panel. The SEI characteristic EoS results are shown by filled squares in the four panels, whereas the results for  SEI EoSs other than L=L$_C$ are in black filled circles. The results of other models are displayed by symbols as mentioned in the legends \cite{Agbemava2014,HFB_masstable}.}
		\label{34Ar34S_36Ca36S_38Ca38Ar_54Ni54Fe_g42SEIG}
	\end{center}
\end{figure*}
 We show in Table \ref{tab-drch} the existing experimental data on
$\Delta$R$_{CH}$ for the $^{34}$Ar-$^{34}$S, $^{36}$Ca-$^{36}$S,
$^{38}$Ca-$^{38}$Ar and $^{54}$Ni-$^{54}$Fe mirror pairs \cite{Brown2017,Brown2020,Pineda2021} along with the predictions computed with the SEI characteristic EoS (L=L$_C$=76.71 MeV).
These SEI results for $\Delta$R$_{CH}$ of the four mirror pairs are also
displayed by filled black squares in the four panels of Fig.\ref{34Ar34S_36Ca36S_38Ca38Ar_54Ni54Fe_g42SEIG} together with the experimental data 
and the predictions of several nonrelativistic (NR) and covariant EDFs~\cite{Agbemava2014,HFB_masstable}.
We see from Table \ref{tab-drch} that the SEI predictions with the characteristic $L_C$ value
successfully remain well within the experimental range for the charge radii differences
in the three pairs $^{34}$Ar-$^{34}$S, $^{36}$Ca-$^{36}$S and $^{54}$Ni-$^{54}$Fe,
and that in the $^{38}$Ca-$^{38}$Ar pair its $\Delta R_{CH}$ value lies just at the upper boundary of experiment.
As can be realized from the results for several NR and covariant models 
displayed in Fig.\ref{34Ar34S_36Ca36S_38Ca38Ar_54Ni54Fe_g42SEIG},
this degree of agreement with the experimental values in these four mirror pairs 
seems to be difficult to satisfy by other mean field interactions.

At the mean-field level the nuclei of the mirror pairs $^{34}$Ar-$^{34}$S, $^{36}$Ca-$^{36}$S,
$^{38}$Ca-$^{38}$Ar and $^{54}$Ni-$^{54}$Fe computed with effective interactions
such as the Gogny D1S and the Skyrme SLy4, SkP and SkM* forces show a soft potential energy surface,
which is compatible with a spherical shape. Dynamical quadrupole deformations can be estimated by
performing beyond mean-field calculations. Consistent CHFB+5DCH calculations 
with the D1S interaction by Delaroche et al.~\cite{Delaroche2010} reveal that the change
in the charge radius due to dynamical deformations is at most 4\% in this interaction.
For the mirror pairs analyzed here, it is seen \cite{Delaroche2010} that the variation of the charge radius
of each partner is almost identical and therefore the impact of dynamical deformation on the 
difference of the charge radii in a mirror pair is negligible. 
As far as the deformation properties of SEI are quite similar to the ones of D1S 
\cite{Behera2016}, we expect that beyond mean-field calculations with SEI would have a 
minimal effect on the charge radii differences obtained for the mirror pairs
in our present spherical calculations.
\begin{figure*}[t]
	\includegraphics[width=0.80\textwidth]{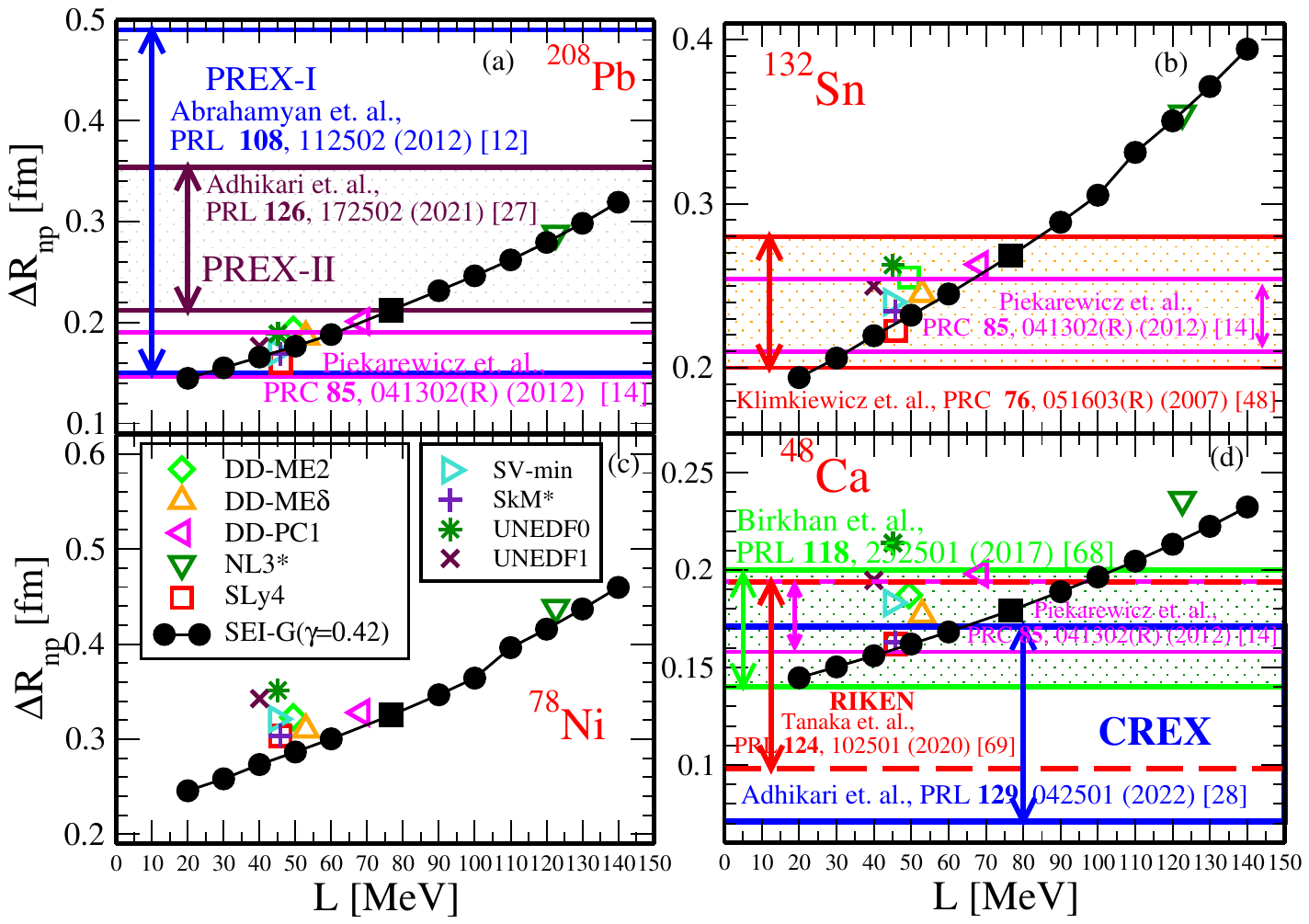}
	\caption{Neutron skin thickness $\Delta R_{np}$ as a function of L for (a) $^{208}Pb$, (b) $^{132}Sn$, (c) $^{78}Ni$, and (d) $^{48}Ca$ for SEI-G($\gamma=0.42$). The results extracted from theoretical analyses of the experiments are shown in horizontal bands in each panel where available.}
	\label{Pb_Sn_Ni_Ca_Rskin}
\end{figure*}

The dependence of $\Delta$R$_{CH}$ 
of mirror pairs on the symmetry energy slope L
has been studied within their experimental uncertainties using NR and covariant EDFs covering a wide-range of L-values
\cite{Brown2017,Brown2020,Pineda2021,Yang2018} in order to ascertain the admissible range of values for L.
To estimate the range of the slope parameter L of the EoS for which the predicted $\Delta R_{CH}$ is within the experimental error bar, 
we have computed $\Delta$R$_{CH}$ for these four mirror pairs by varying the L-value of the SEI sets. As we mentioned in Sec.~II, under the procedure of variation of L in SEI the binding energies and {\it {rms}} mass radii of
nuclei remain very stable. For example, for the present nuclei in the range of
variation of L from 20 to 110 MeV the BEs of $^{34,38}$Ar, $^{34}$S, $^{38}$Ca, $^{54}$Ni, and 
$^{54}$Fe change by less than 0.5\%, and in $^{36}$Ca and $^{36}$S this change is $\approx$1\%.
The $\Delta$R$_{CH}$ values obtained under the variation of L
are shown by filled black circles in the four panels of Fig.\ref{34Ar34S_36Ca36S_38Ca38Ar_54Ni54Fe_g42SEIG}. 
The upper limit of the slope L of the symmetry energy for which the experimental error bars of these four pairs are covered by the SEI 
EoS comes out to be L$\approx$100 MeV.

If we compare to the PREX-2 data on the neutron skin $\Delta$R$_{np}$ in $^{208}$Pb, shown in panel (a) of 
Fig.\ref{Pb_Sn_Ni_Ca_Rskin}, we see that SEI with its characteristic L$_C$ predicts a neutron skin that coincides with the lower bound of the PREX-2 
range. In order to reproduce the central value of 0.283 fm, the SEI requires a value of L=122 MeV.
In the same Fig.\ref{Pb_Sn_Ni_Ca_Rskin} we also display the skin 
thickness predicted by SEI as a function of L in $^{132}$Sn, $^{78}$Ni and $^{48}$Ca together with the results (except for $^{78}$Ni where no data 
are available) obtained from the different theoretical analyses of experimental data. 
In the panel for $^{132}$Sn, the data extracted from the pygmy dipole resonance study of \cite{Klimkiewicz2007} is shown.
The very recent CREX data \cite{Adhikari2022} on the
$^{48}$Ca skin thickness, $\Delta R{_{np}^{48}}$=0.121$\pm$0.026(exp)$\pm0.024$(model) fm=0.071 - 0.171 fm, shown in panel (d) of this figure, demands L$\le$65 MeV for the SEI EoS in order to remain within the CREX uncertainty limit. However, in order to predict the central value 0.121 fm, SEI requires a very small value for L.
 The predictions of the SEI EoSs have a larger overlap with the results on $\Delta$R$_{np}$ of $^{48}$Ca
obtained from the dipole polarizability study \cite{Birkhan2017} and from the scattering of $^{48}$Ca on C at 280 MeV/nucleon 
\cite{Tanaka2020}, which are displayed in the same panel (d) of Fig.\ref{Pb_Sn_Ni_Ca_Rskin}.
The correlation between $\Delta$R$_{np}$ and the proton radii difference in mirror pairs, Eqs.(\ref{Eq_R_np}) and (\ref{Eq_ab}), could be the
yardstick to these various measurements in medium and heavy nuclei, but nature forbids by not providing their mirror pair. 
%

\begin{figure*}[!tbp]
	\begin{center}
		\includegraphics[width=0.80\textwidth]{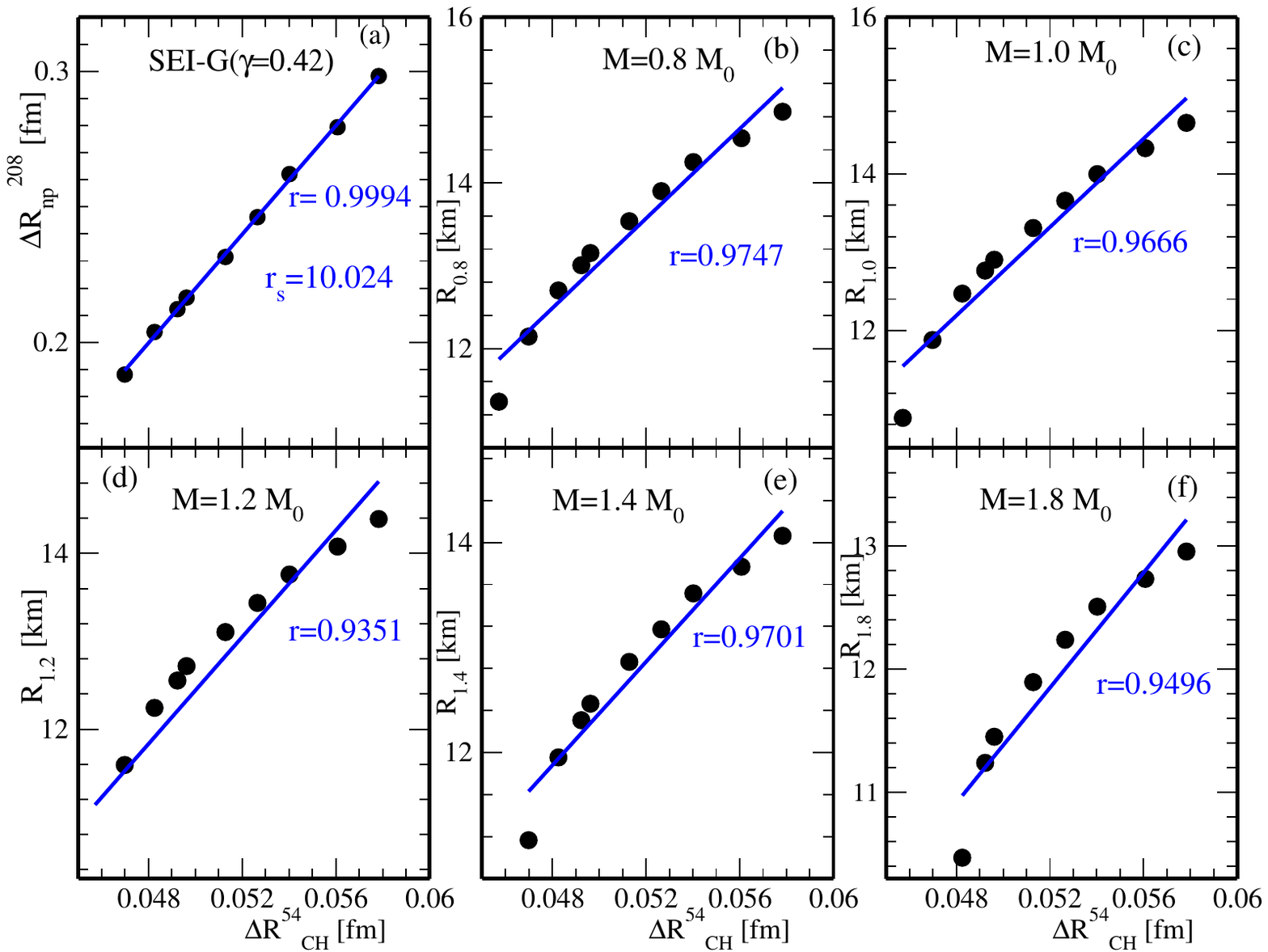}
		\caption{Stellar radii for neutron stars having masses (b) M=0.8 M$_{\odot}$, (c) 1.0 M$_{\odot}$, (d) 1.2 M$_{\odot}$, (e) 1.4 M$_{\odot}$, and (f) 
			1.8 M$_{\odot}$ as a function of the charge radii difference $\Delta R_{CH}^{54}$ in the $^{54}$Ni-$^{54}$Fe mirror pair. The
			correlation coefficient r obtained from linear regression is given in each panel. The correlation between $\Delta$R$_{CH}^{54}$ 
			and $\Delta$R$_{np}^{208}$ 
			is shown in panel (a).}
		\label{R08112141618_54Rch}
	\end{center}
\end{figure*}

From the  correlation between $\Delta$R$_{CH}$ and L shown in Fig.\ref{34Ar34S_36Ca36S_38Ca38Ar_54Ni54Fe_g42SEIG}, we see that 
$\Delta$R$_{CH}$ increases with L. 
This increase in $\Delta$R$_{CH}$ for increasing L stems from the combined effect of Coulomb force and neutron pressure in nuclei. 
The pressure
in pure neutron matter at saturation density $\rho_0$ is related to L as
\begin{equation}
P_{N}(\rho_{0})\approx\frac{\rho_{0} L}{3}.
\label{Eq_pnm}
\end{equation}
On account of the Coulomb force, regardless of the L-value of the EoS, the proton distribution is pushed out compared to the neutron distribution. 
Due to this, and taking into account the fit in Eqs.(\ref{Eq_R_np}) and (\ref{Eq_ab}), we realize that the neutron skin $\Delta$R$_{np}$ has a small negative 
value when $\Delta$R$_{CH}$=0. Thus, proton-rich nuclei in the isotopic and isotonic chains displayed in 
Fig. \ref{N14_N28_N50_Z10_Z20_Z28_g42SEIG} have a negative neutron skin when computed with SEI.
On the other hand, the nuclear
surface is pushed out as L grows, owing to the increase in neutron pressure, as can 
be seen from Eq.(\ref{Eq_pnm}), resulting in an extension of the neutron (proton) distribution at the surface in neutron (proton) rich nuclei. Whether 
it is a finite nucleus or a neutron star, both systems are governed by the same strong interactions and the EoS 
relating pressure to density \cite{Thiel2019,Novario2020,Shen2020,Horowitz2019,Wei2020}; in a nucleus the neutron pressure acts against the nuclear 
surface tension, whereas in a NS it acts against gravity.
Hence, the neutron skin thickness and the NS radius have a correspondence 
in the context of the L-dependence, in spite of their very different size.

The correlation between the skin thickness $\Delta$R$_{np}^{208}$ in the $^{208}$Pb-nucleus and the radius of neutron stars was 
investigated in Refs.\cite{Horowitz2001,Carrier2003}. SEI calculations show that the charge radii difference $\Delta$R$_{CH}^{54}$
in the mirror pair $^{54}$Ni-$^{54}$Fe is strongly correlated to $\Delta$R$_{np}^{208}$, as can be seen in panel (a) of 
Fig.\ref{R08112141618_54Rch}, with correlation coefficient 0.9994 and regression slope r$_S$=10.02. 
 Due to the high accuracy reached in the measurement of $\Delta$R$_{CH}^{54}$ \cite{Pineda2021}, this quantity seems an interesting 
candidate for correlating with different properties of stellar objects as an alternative to the skin thickness of $^{208}$Pb, which 
presently has a larger uncertainty depending on the nature of the experiment performed to determine it.
Keeping this in view, 
 we have analyzed the behaviour of the stellar radius R$_{NS}$ of NSs having masses from 0.8 M$_{\odot}$ to 1.8 M$_{\odot}$ as a
function of $\Delta$R$_{CH}^{54}$, obtained from the SEI EoSs with different slope parameter L. The results are shown in panels (b)-(f) 
of Fig.\ref{R08112141618_54Rch}. 
In solving the TOV equations we have used the BPS-BBP EOS \cite{Baym1971a,Baym1971b} up to a density 0.07468 $fm^{-3}$
and the SEI EOS for densities thereafter.
In each of these NSs of different masses, we find a linear correlation between the radius R$_{NS}$ and the charge radii difference 
$\Delta$R$_{CH}^{54}$. When the mass of the NS grows, the quality of the linear fits shows a slight decreasing trend,
which was first pointed out by Yang and Piekarewicz in Ref.\cite{Yang2018}. In that work the correlation between the radius of
low-mass NSs and the difference of proton radii in the mirror pair $^{50}$Ni-$^{50}$Ti was analyzed using well calibrated
covariant EDFs covering a range of L$\approx$50-140 MeV.
The argument given in  \cite{Yang2018} for explaining this decreasing trend was that the slope parameter determines the EoS of 
neutron-rich matter around normal NM density and the central density of the low-mass NSs is relatively close to the NM saturation density, 
whereas heavier NSs have larger central densities.
For the present SEI EoSs, the decrease in the quality of the correlation between R$_{NS}$ and $\Delta$R$_{CH}^{54}$, from lighter to heavier NS masses, is found to be relatively moderate.
The  R$_{NS}$-$\Delta R_{CH}^{54}$ 
correlation in the 1.4 M$_{\odot}$ NS, displayed in panel (e) of Fig.\ref{R08112141618_54Rch}, 
together with the data in panel (a) of this figure, confirm the correlation between
R$_{1.4}$ and $\Delta$R$_{np}^{208}$ found in Ref.\cite{Reed2021} for a set of well calibrated covariant EDFs.
The NICER data analysis of the signals from the so far measured heaviest-mass pulsar PSR J0740+6620 \cite{Miller2021}, prescribes a limit on the 
radius of a 1.4 M$_{\odot}$ NS to be R$_{1.4}$=12.45$\pm$0.65 km at 1$\sigma$-level. Another different range, R$_{1.4}$=11.9$\pm$1.4 
km, was ascertained from the data analysis of the GW170817 event of a binary NS merger in Ref.\cite{Abbott2018} by the LIGO/VIRGO Collaboration. 
 This result prescribes the upper limit to be 13.3 km, which is close to the 13.1 km value of the PSR J0740+6620 data. However, 
there is larger uncertainty in the lower limiting value, which is 10.5 km from the GW170817 data in comparison to 11.8 km from 
NICER PSR J0740+6620 data. From the R$_{NS}$-$\Delta R_{CH}^{54}$ correlation
of Fig.\ref{R08112141618_54Rch}-(e) together with the  $\Delta R_{CH}$-L correlation in Fig.\ref{34Ar34S_36Ca36S_38Ca38Ar_54Ni54Fe_g42SEIG}-(d), we constrain the slope parameter L in the range $\approx$ 70-100 MeV from the NICER 1$\sigma$-level R$_{1.4}$ data of PSR J0740+6620.
This range for L, though larger than the values extracted from theoretical studies \cite{Hebeler2013,Drischler2020,Steiner2012,Gandolfi2014} and from
theoretical analyses of various experimental data \cite{Zhang2013,Hagen2016,Chen2010,Roca2015,Essick2021}, conforms to the range 106$\pm$37 MeV 
obtained in Ref.\cite{Reed2021} from the analysis of the PREX-2 data together with the L-$\Delta$R$^{208}_{np}$ correlation using a set of
covariant EDFs. Also it is in good agreement with the range 70$\le$$L$$\le$101 
MeV extracted from the nucleon-nucleus elastic scattering cross-section analyses of reactions involving isobaric analog states 
\cite{Danielewicz2017}. Our result also falls well within the range 42$\le$$L$$\le$117 MeV estimated 
from the charged pion ratio measurement at high 
transverse momentum in neutron-rich Sn+Sn collisions \cite{Estee2021}, as well as within the range 63$\le$$L$$\le$113 MeV ascertained from the transport model analysis of the isospin diffusion data~\cite{chen2005}. 

\begin{figure*}[t]
	\begin{center}
		\includegraphics[width=0.80\textwidth]{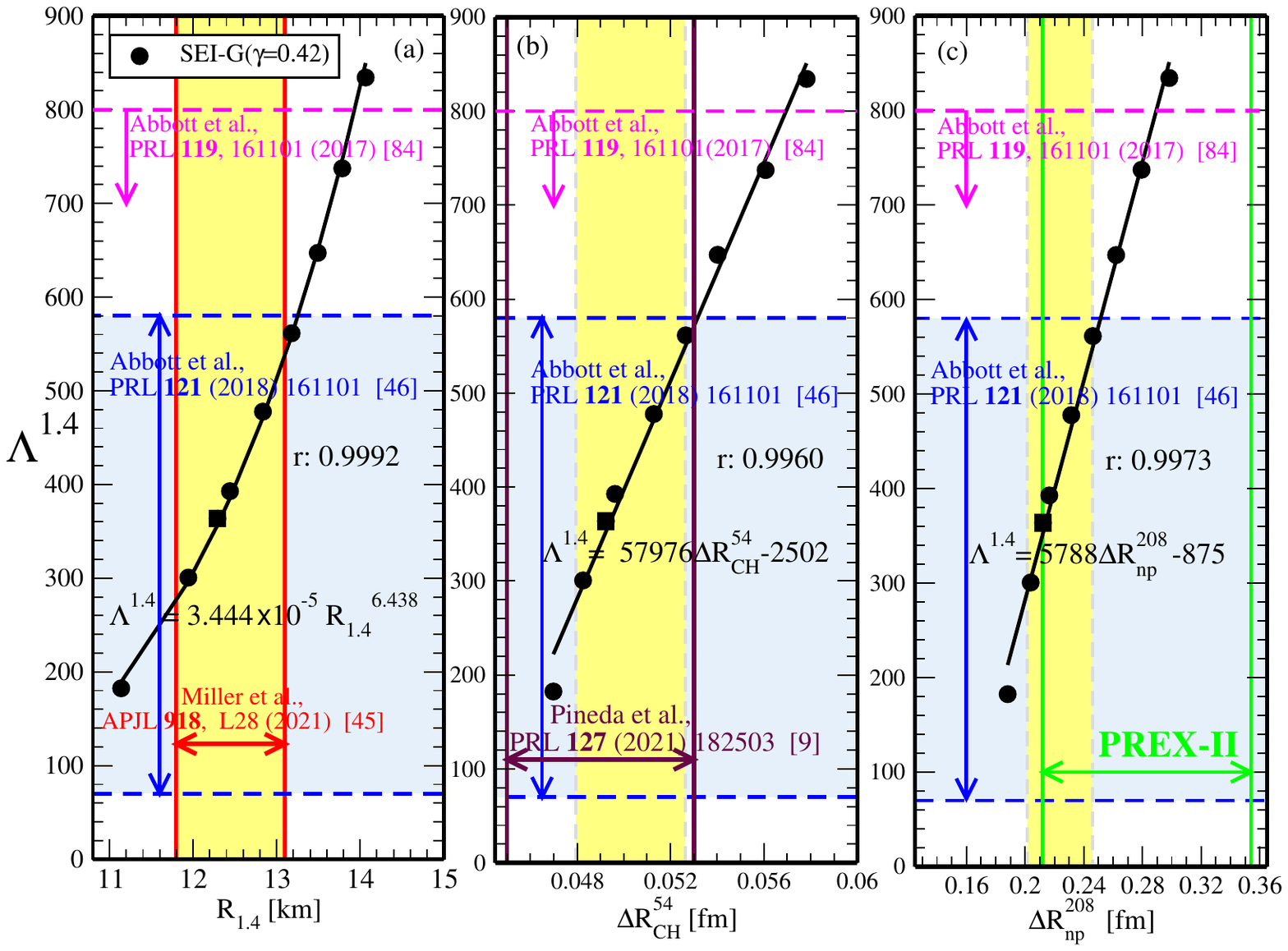}
		\caption{Predictions of SEI-G($\gamma$=$0.42$) for the tidal deformability, $\Lambda^{1.4}$, in 1.4M$_\odot$ NSs shown by 
filled black circles versus (a) the radius $R_{1.4}$ of 1.4M$_\odot$ NSs, (b) charge radii difference $\Delta$R$^{54}_{CH}$ of the 
$^{54}$Ni-$^{54}$Fe mirror pair, and (c) neutron skin thickness $\Delta$R$_{np}^{208}$ in $^{208}$Pb. 
 The black squares correspond to the SEI calculation with characteristic slope parameter L$_C$=76.71 MeV.
The GW170817 2$\sigma$-limits on 
$\Lambda^{1.4}$ \cite{Abbott2018} are shown between the blue horizontal lines and the earlier upper limit \cite{Abbott2017} by the pink line. 
In panel (a), the 1$\sigma$-limits on R$_{1.4}$ from pulsar PSR J0740+6620 data \cite{Miller2021} are shown by red vertical lines. 
In (b) the experimental data for $\Delta$R$^{54}_{CH}$ \cite{Pineda2021} are shown between maroon lines, and in (c) the PREX-2 limits 
\cite{Adhikari2021} are shown by green vertical lines. In each panel the vertical yellow bands correspond to the SEI values in the range of L=70-100 MeV for the respective variables $R_{1.4}$, $\Delta$R$^{54}_{CH}$, and~$\Delta$R$_{np}^{208}$.}
		\label{Tidal_Rskin208_R14}
	\end{center}
\end{figure*}

The dimensionless tidal deformability parameter, $\Lambda^{1.4}$, for 1.4 M$_{\odot}$ NSs
computed with SEI in the range L=70--100 MeV 
is found to be compatible with the range $\Lambda^{1.4}$
=190$^{+390}_{-120}$ constrained from the analysis of the GW170817 data at 90\% confidence 2$\sigma$-level \cite{Abbott2018}, which updates the previous upper limit 
of $\le$800 \cite{Abbott2017}. In panels (a) and (b) of Fig.\ref{Tidal_Rskin208_R14} we show by black filled circles the tidal 
deformability $\Lambda^{1.4}$ computed with SEI versus R$_{1.4}$ and $\Delta$R$_{CH}^{54}$, respectively. Along the horizontal axis, in both 
panels, the R$_{1.4}$ and $\Delta$R$_{CH}^{54}$ values corresponding to the range of L=70-100 MeV are shown by the vertical yellow bands.
 The $\Lambda^{1.4}$ values predicted by SEI within L=70-100  MeV lie in a range of 300--560,
which is within the 
constraint extracted from GW170817 \cite{Abbott2018} (shown between two horizontal lines in blue).
 We note that the NICER data for R$_{1.4}$ are at 1$\sigma$ confidence, whereas the LIGO $\Lambda^{1.4}$ data are at 
2$\sigma$-level accuracy. Thus, the inferences with regards to the $\Lambda^{1.4}$ constraint of GW170817 are subject to correction on the 
future availability of its data at 1$\sigma$ confidence.
We also see from panel (a) of Fig.\ref{Tidal_Rskin208_R14} a strong power-law correlation between $\Lambda^{1.4}$ computed with SEI and R$_{1.4}$,  which scales as 
(R$_{1.4})^{6.4}$ (\cite{Lourenso2020} and references therein).

%

In order to estimate the degree of agreement of the SEI predictions with the PREX-2 data \cite{Adhikari2021}, we show in panel (c) of 
Fig.\ref{Tidal_Rskin208_R14} the tidal deformability $\Lambda^{1.4}$ against the neutron skin in $^{208}$Pb. The SEI values of 
$\Delta$R$_{np}^{208}$=0.203-0.246 fm in the range L=70-100 MeV are shown by the vertical yellow band.
and the PREX-2 data by 
two vertical green lines. As expected, a strong linear correlation between $\Lambda^{1.4}$ and $\Delta$R$_{np}^{208}$ is found. 
Although the upper limiting value L=100 MeV can reproduce the PREX-2 data up to $\Delta$R$_{np}^{208}$ $\approx$ 0.25 fm, it fails
in reproducing its central value 0.283 fm. However, $\Delta$R$_{np}^{208}$ predicted by SEI is compatible with 
the range 0.18 $\pm$ 0.07 fm found from the dispersive optical model 
analysis \cite{Washington2020}. If we take 
as upper limit of R$_{1.4}$ the value 14.26 km from the NICER PSR J0030+0451 data \cite{Miller2019}, 
the PREX-2 result up to $\approx$ 0.31 fm can be explained, but in that case the tidal deformability constraint will be violated.
 The lower limiting value of the slope parameter L=70 MeV satisfies the $\Lambda^{1.4}$ constraint and predicts
$\Delta$R$_{np}^{208}$=0.20 fm somewhat below the PREX-2 lower bound ($\Delta$R$_{np}^{208}$=0.212 fm). 
This PREX-2 lower bound is predicted by the SEI EoS with the characteristic slope parameter L$_C$=76.71 MeV.
The recent CREX result for the neutron skin of $^{48}$Ca, $\Delta$R$_{np}^{48}$= 0.12$\pm$0.026(exp)$\pm$0.024(model) fm \cite{Adhikari2022},
indicates a softer EoS. Our calculations using SEI with L between 70 and 100 MeV
predict $\Delta$R$_{np}^{48}$ in the range 0.175-0.195~fm, which is consistent with the model-averaged result 
$\Delta$R$_{np}^{48}$ =0.176 $\pm$0.018 fm obtained in the EDF study in Ref.\cite{Piekarewicz2012}.

To examine the possibility of reproducing the CREX and PREX-2 data simultaneously within the present functional form of the SEI interaction,
we have varied the basic nuclear matter properties of SEI without reference to the predictions of the EoS of $\gamma$=0.42.
Although a certain effect from the n-p effective mass splitting of SEI is found, the symmetry energy and its slope parameter have,
as expected, the largest influence on the variation of the neutron skin thickness.
By decreasing $E_s(\rho_0)$ and L$(\rho_0)$ to $\approx$30 MeV and $\approx$20 MeV, respectively, the central value of the CREX data can be reproduced,
but in that case $\Delta$R$_{np}^{208}$ gets substantially underestimated. If we increase $E_s(\rho_0)$ and L$(\rho_0)$ to reproduce the
PREX-2 data, then $\Delta$R$_{np}^{48}$ gets much overestimated compared to CREX.
The situation is reminiscent of the findings of the recent studies reported in
Refs.\cite{Reinhard2022,Paar2023}. In these works it has been concluded that it is not possible to reproduce the CREX and PREX-2 data 
simultaneously with the present form of the nuclear EDFs. A similar situation is encountered in the chiral EFT calculations, where 
the {\it ab initio} theory predicts $\Delta$R$_{np}^{48}$ in the range of CREX \cite{Hagen2016}, whereas the {\it ab initio} approach combined 
with recent advances in many-body techniques predicts $\Delta$R$_{np}^{208}$ in the range 0.12-0.22 fm \cite{Hu2022}.
This is consistent with the L-value of the chiral EFT calculations, which lies in the range L$\approx$40-60 MeV.
In a separate work \cite{Tagami2022}, the L-$\Delta$R$_{np}^{48}$ and L-$\Delta$R$_{np}^{208}$ correlations have been analyzed using 207 EDFs.
The authors of \cite{Tagami2022} find that two non-overlapping ranges for L are required in order to predict $\Delta$R$_{np}^{48}$
and $\Delta$R$_{np}^{208}$ within the CREX and PREX-2 ranges, in agreement with the conclusions of Refs.\cite{Reinhard2022,Paar2023}.
 
\section{Conclusions}
 The isospin-symmetry breaking effect leading to a linear correlation between the proton {\it{rms}} radii difference in mirror pairs
and neutron skin thickness in nuclei, instead of equality between them, earlier found in the context of low-energy chiral
EFT, Skyrme, and covariant functionals, is confirmed by the calculations with the SEI finite-range model.
 The SEI model allows one to generate different parameter sets differing in their predictions in the 
isovector sector, which are characterized by the slope L of the symmetry energy, whereas the isoscalar properties
 remain invariant. We find that the available experimental data for charge radii differences in mirror nuclei pairs constrain L to the range
$\le$100 MeV.  On the basis of the correlation found between the mirror pair charge radii difference and the NS radius, we have 
analyzed the radius of a NS of 1.4$M_{\odot}$ as a function of the charge radii difference in the mirror nuclei $^{54}$Ni 
and $^{54}$Fe using the SEI model, since the $\Delta$R$^{54}_{CH}$ observable can be experimentally determined to high accuracy \cite{Pineda2021}. 
The L$\le$100 MeV limit extracted from the data of the charge radii 
differences in mirror nuclei pairs is
found to be consistent with the R$_{1.4}$=12.45$\pm$0.65 km radius constraint of the NICER data analysis for the pulsar 
PSR J0740+6620, which further constrains the slope parameter L in the range 70-100 MeV.
In this range of $L$, the tidal deformability $\Lambda^{1.4}$=190$^{+390}_{-120}$ extracted from the
GW170817 event at 2$\sigma$ level is well reproduced.
In addition, the SEI model with L between 70 and 100 MeV explains the PREX-2 data
on $\Delta$R$_{np}^{208}$ up to $\approx$ 30\% of the experimental range.
 But the CREX result for $\Delta$R$_{np}^{48}$ is not reproduced,
providing with L=70 MeV a theoretical estimate at the upper limit $\approx$0.17 fm of the CREX value.
 This might be attributed to the inherent limitations
in the functional form of the present NR and covariant EDFs discussed in Refs.\cite{Reinhard2022,Paar2023}. 
In the context of PVES experiments,
more precise electroweak measurements of the skin thickness of neutron-rich nuclei
will be boosted by full operation of the future Mainz Energy-recovery Superconducting Accelerator (MESA)~\cite{MESA2023}.
On the other hand, the continuing advances in laser spectroscopy may provide 
more data on charge radii along chains of mirror pairs, which  
can contribute to reducing the model dependence of the theoretical analyses and lead to new
constraints on the properties of the isospin-dependent EoS.

\section*{Acknowledgements}
P.B. acknowledges support from MANF Fellowship of UGC, India. T.R.R. offers sincere thanks to Prof. B. Behera for 
useful discussions. M.C. and X.V. were partially supported by Grants No.\ PID2020-118758GB-I00 and No.\ CEX2019-000918-M
(through the ``Unit of Excellence Mar\'{\i}a de Maeztu 2020-2023'' award to ICCUB)
from the Spanish MCIN/AEI/10.13039/501100011033.


\begin{thebibliography}{99}
	\bibitem{Yang2023}
	X. F. Yang et al., Prog. Part. Nucl. Phys. \textbf{129}, 104005 (2023).
	\bibitem{Campbell2016} 
	P. Campbell, I. Moore, and M. Pearson, Prog. Part. Nucl. Phys. \textbf{86}, 127 (2016).
	\bibitem{Malbrunot-Ettenauer2022} 
	S. Malbrunot-Ettenauer, S. Kaufmann, S. Bacca, C. Barbieri, J. Billowes, et al., Phys. Rev. Lett. \textbf{128}, 022502 (2022).
	\bibitem{Brown2017}
	B. A. Brown, Phys. Rev. Lett. \textbf{119}, 122502 (2017).
	\bibitem{Yang2018}
	J. Yang and J. Piekarewicz, Phys. Rev. C \textbf{97}, 014314 (2018).
	\bibitem{Sammarruca2018}
	F. Sammarruca, Front. Phys., \textbf{6}, 90 (2018).
	\bibitem{Brown2020}
	B. A. Brown, K. Minamisono, J. Piekarewicz, et al., Phys. Rev. Research \textbf{2}, 022035(R) (2020).
	%
	\bibitem{Gaidarov2020}
	M. Gaidarov, I. Moumene, et al., Nucl. Phys. A \textbf{1004}, 122061 (2020).
	%
	\bibitem{Pineda2021}
	S. V. Pineda, K. K\"onig, D. M. Rossi, B. A. Brown, et al., Phys. Rev. Lett. \textbf{127}, 182503 (2021).
	%
	\bibitem{ReinhardL2022}
	P.-G. Reinhard and W. Nazarewicz, Phys. Rev. C \textbf{105}, L021301 (2022).
	%
	\bibitem{Novario2023}
	S.J. Novario, D. Lonardoni, S. Gandolfi, and G. Hagen, Phys. Rev. Lett. \textbf{130}, 032501 (2023).
	\bibitem{Abrahamyan2012}
	S. Abrahamyan, Z. Ahmed, et al., Phys. Rev. Lett. \textbf{108}, 112502 (2012).
	\bibitem{Tamii2011}
	A. Tamii, I. Poltoratska, et al., Phys. Rev. Lett. \textbf{107},  062502 (2011).
	\bibitem{Piekarewicz2012}
	J. Piekarewicz, B. K. Agrawal, et al., Phys. Rev. C \textbf{85}, 041302(R) (2012).
	\bibitem{Rossi2013}
	D. M. Rossi, P. Adrich, et al., Phys. Rev. Lett. \textbf{111}, 242503 (2013).
	%
	\bibitem{Brown2000}
	B. A. Brown, Phys. Rev. Lett. \textbf{85}, 5296 (2000).
	
	\bibitem{Reinhard2010}
	P.-G. Reinhard and W. Nazarewicz, Phys. Rev. C \textbf{81}, 051303(R) (2010).
	%
	\bibitem{Thiel2019}
	M. Thiel, C. Sfienti, J. Piekarewicz, C. J. Horowitz, and M. Vanderhaeghen, J. Phys. G \textbf{46}, 093003 (2019).
	%
	%
	\bibitem{Piekarewicz2019}
	J. Piekarewicz and F. J. Fattoyev, Physics Today \textbf{72}, 7, 30 (2019).
	%
	%
	\bibitem{Fattoyev2018}
	F. J. Fattoyev, J. Piekarewicz and C. J. Horowitz, Phys. Rev. Lett. \textbf{120}, 172702 (2018).
	%
	\bibitem{Reed2021}
	B T Reed, F. J. Fattoyev, et al., Phys. Rev. Lett. \textbf{126}, 172503 (2021).
	%
	\bibitem{Zhang2020}
	Y. Zhang, M. Liu, C.-J. Xia, Z. Li, and S. K. Biswal, Phys. Rev. C
\textbf{101}, 034303 (2020).
\bibitem{Guven2020}
H. G\"uven, K. Bozkurt, E. Khan, and J. Margueron, Phys. Rev. C \textbf{102}, 015805 (2020).
%
\bibitem{Baiotti2019}
 L. Baiotti, Prog. Part. Nucl. Phys. \textbf{109}, 103714 (2019).
%
\bibitem{Tsang2019PLB}
M. Tsang, W. Lynch, P. Danielewicz, and C. Tsang, Phys. Lett. B \textbf{795}, 533 (2019).
%
\bibitem{Fasano2019}
M. Fasano, T. Abdelsalhin, A. Maselli, and V. Ferrari, Phys. Rev. Lett. \textbf{123}, 141101 (2019).
	
	\bibitem{Adhikari2021}
	D. Adhikari et al., Phys. Rev. Lett. \textbf{126}, 172502 (2021).
	\bibitem{Adhikari2022}
	D. Adhikari , H. Albataineh et al., Phys Rev Lett., \textbf{129}, 042501 (2022).
	%
	\bibitem{Hagen2016}
	G. Hagen, A. Ekstr\"om, C. Forss\'en, G. R. Jansen, W. Nazarewicz, et al., Nat. Phys. \textbf{12}, 186 (2016).
	%
	\bibitem{simonis2019}
	J. Simonis, S. Bacca, G. Hagen, Eur. Phys. J. A \textbf{55}, 241 (2019).
	%
	\bibitem{Paar2023}
	E. Y\"{u}ksel and Nils Paar, Phys. Lett. B \textbf{836}, 137622 (2023).
	%
	\bibitem{Tagami2022}
	S. Tagami, T. Wakasa, and M. Yahiro, Results in Phys. \textbf{43}, 106037 (2022).
	%
	\bibitem{Reinhard2022}
	P.-G. Reinhard, X. Roca-Maza, W. Nazarewicz, Phys. Rev. Lett. \textbf{129}, 232501 (2022).
	%
	\bibitem{Hu2022}
  B. Hu, W. Jiang, T. Miyagi, et al., Nat. Phys. \textbf{18}, 1196 (2022).  
	%
	\bibitem{Behera1998}
	B Behera, T R Routray and R K Satpathy, J. Phys. G: Nucl. Part. Phys. \textbf{24}, 2073 (1998).
	%
	\bibitem{Behera2013}
	B Behera, X Vi\~nas, et al., J. Phys. G: Nucl. Part. Phys. \textbf{40}, 095105 (2013).
	\bibitem{Behera2015}
	B Behera, X Vi\~nas, T R Routray and M Centelles, J. Phys. G: Nucl. Part. Phys. \textbf{42}, 045103 (2015).
	\bibitem{Routray2021}
	T. R. Routray, P. Bano et al., Phys. Rev. C, \textbf{104}, L011302 (2021).	
	\bibitem{Routray2022}
	P. Bano, X. Vi\~nas, T. R. Routray, M. Centelles, M. Anguiano, and L. M. Robledo, Phys. Rev C, \textbf{106}, 024313 (2022).
	%
	\bibitem{trr2007}
	B. Behera, T.R. Routray, A. Pradhana, S.K. Patra, and P.K. Sahu, Nucl. Phys. A\textbf{794}, 132 (2007).
	%
	\bibitem{trr2016}
	T. R. Routray, X. Vi\~{n}as, D. N. Basu, S. P. Pattnaik, M. Centelles, L. M. Robledo, and B. Behera, J. Phys. G: Nucl. Part. Phys. \textbf{43}, 105101 (2016).
	%
	\bibitem{trr2018}
	S. P. Pattnaik, T. R. Routray, X. Vi\~{n}as, D. N. Basu, M. Centelles, K. Madhuri, and B. Behera, J. Phys. G: Nucl. Part. Phys. \textbf{45}, 055202 (2018).
	%
	\bibitem{mario2019}
	C. Gonzalez-Boquera, M. Centelles, X. Vi\~{n}as, and T. R. Routray, Phys. Rev. C \textbf{100}, 015806 (2019).
	%
	\bibitem{trr2021}
	T. R. Routray, S. P. Pattnaik, C. Gonzalez-Boquera, X.Vi\~{n}as, M.Centelles and B. Behera, Phys. Scr. \textbf{96}, 045301 (2021).	
	%
	\bibitem{Miller2021}
	M. C. Miller, F. K. Lamb, et al., Astrophys. J. Lett. \textbf{918}, L28 (2021). 
	\bibitem{Abbott2018}
	B. P. Abbott et al. (Virgo and LIGO Scientific Collaborations), Phys. Rev. Lett. \textbf{121}, 161101 (2018).
	%
	\bibitem{Roca2015}
	X. Roca-Maza, X. Vi\~{n}as, M. Centelles, et al., Phys. Rev. C \textbf{92}, 064304 (2015).
	%
	\bibitem{Klimkiewicz2007}
	A. Klimkiewicz, N. Paar et al., Phys. Rev. C \textbf{76}, 051603(R) (2007).
	%
	\bibitem{Drischler2020}
	C. Drischler, R. J. Furnstahl, J. A. Melendez, and D. R. Phillips, Phys. Rev. Lett. \textbf{125}, 202702 (2020).
	%
	\bibitem{Essick2021}
  R. Essick, I. Tews, P. Landry, and A. Schwenk, Phys. Rev. Lett. \textbf{127}, 192701 (2021).
	%
	\bibitem{Tsang2019}
	C. Y. Tsang, B. A. Brown, et al., Phys. Rev. C \textbf{100}, 062801(R) (2019).
	%
	\bibitem{Danielewicz2017}
	P. Danielewicz, P. Singh, and J. Lee, Nucl. Phys. A \textbf{958}, 147 (2017).
	\bibitem{Estee2021}
	J. Estee, W. G. Lynch et al., Phys. Rev. Lett. \textbf{126}, 162701 (2021).
	%
	\bibitem{Perera2021}
	U. C. Perera, A. V. Afanasjev, and P. Ring, Phys. Rev. C, \textbf{104}, 064313 (2021).
	%
	\bibitem{ong2010}
        A. Ong, J. C. Berengut, and V. V. Flambaum, Phys. Rev. C \textbf{82}, 014320 (2010).
	%
	\bibitem{Behera09}
	B. Behera, T. .R Routray, and S. K. Tripathy, J. Phys. G: Nucl. Part. Phys. \textbf{36}, 125105 (2009).
	%
	\bibitem{Sammarruca2010}
	F. Sammarruca, Int. J. Mod. Phys. E \textbf{19}, 1259 (2010).
	\bibitem{Wiringa1988}
	R. B. Wiringa, Phys. Rev. C \textbf{38}, 2967 (1988).
	\bibitem{APR1998}
	A. Akmal, V. R. Pandharipande, and D. G. Ravenhall, Phys. Rev. C \textbf{58}, 1804 (1998).
	\bibitem{Soubbotin2000}
	V. B. Soubbotin and X. Vi\~nas, Nucl. Phys. A \textbf{665}, 291 (2000).
	\bibitem{Soubbotin2003}
	V. B. Soubbotin, V. I. Tselyaev, and X. Vi\~nas, Phys. Rev. C \textbf{67}, 014324 (2003).
	\bibitem{Estal2001}
	M. Del Estal, M. Centelles, X. Vi\~nas and S. K. Patra, Phys. Rev. C \textbf{63}, 044321 (2001).
	\bibitem{Bertsch1991}
	G. F. Bertsch and H. Esbensen, Ann. Phys. (N.Y.) \textbf{209}, 327-363 (1991).
	%
	\bibitem{Behera2016}
	B. Behera, X. Vi\~nas, T. R. Routray et al., J. Phys. G: Nucl. Part. Phys. \textbf{43}, 045115 (2016).
	%
	\bibitem{HFB_masstable}
	Mass Explorer: DFT mass tables available at http:// \\
	massexplorer.frib.msu.edu/content/DFTMassTables.html
	%
	\bibitem{Agbemava2014}
	S. E. Agbemava, A. V. Afanasjev, et al., Phys. Rev. C, \textbf{89}, 054320 (2014).
	%
		%
		\bibitem{Delaroche2010}
		J.-P. Delaroche, M. Girod, J. Libert, H. Goutte, S. Hilaire, S. P\'eru, N. Pillet, and G. F. Bertsch,
			Phys. Rev. C, \textbf{81}, 014303 (2010).
	%
	\bibitem{Birkhan2017}
	J. Birkhan, M. Miorelli et al., Phys Rev Lett., \textbf{118}, 252501  (2017). 
	\bibitem{Tanaka2020}
	M. Tanaka et al., Phys. Rev. Lett. \textbf{124}, 102501 (2020).
	%
\bibitem{Novario2020}
S. J. Novario, G. Hagen, G. R. Jansen, and T. Papenbrock, Phys. Rev. C \textbf{102}, 051303(R) (2020).
%
\bibitem{Shen2020}
H. Shen, F. Ji, J. Hu, and K. Sumiyoshi, Astrophys. J. \textbf{891}, 148 (2020).
%
\bibitem{Horowitz2019}
C. Horowitz, Ann. Phys. (Amsterdam) \textbf{411}, 167992 (2019).
%
\bibitem{Wei2020}
J. B. Wei, J. J. Lu, G. F. Burgio, Z. H. Li, and H. J. Schulze, Eur. Phys. J. A \textbf{56}, 63 (2020).
	%
	\bibitem{Horowitz2001}
	C. J. Horowitz and J. Piekarewicz, Phys. Rev. C \textbf{64}, 062802(R) (2001).
	\bibitem{Carrier2003}
	J. Carriere, C. J. Horowitz, and J. Piekarewicz, Astrophys. J. \textbf{593}, 463 (2003).
	\bibitem{Baym1971a}
G. Baym, H. A. Bethe, and C. J. Pethick, Nucl. Phys. A \textbf{175} 225 (1971).
%
\bibitem{Baym1971b}
G. Baym, C. J. Pethick, and P. Sutherland, Astrophys. J \textbf{170} 299 (1971).
	%
	\bibitem{Hebeler2013}
	K. Hebeler, J. Lattimer, C. Pethick, and A. Schwenk,
	Astrophys. J. \textbf{773}, 11 (2013).
	%
	\bibitem{Steiner2012}
	A.W. Steiner and S. Gandolfi, Phys. Rev. Lett. \textbf{108}, 081102  (2012).
	\bibitem{Gandolfi2014}
	S. Gandolfi, J. Carlson, et al., Eur. Phys. J. A \textbf{50}, 10 (2014).
	%
	\bibitem{Zhang2013}
	Z. Zhang and L.-W. Chen, Phys. Lett. B \textbf{726}, 234 (2013).
	\bibitem{Chen2010}
	L.-W. Chen, C. M. Ko, B.-A. Li, and J. Xu, Phys. Rev. C \textbf{82}, 024321 (2010).
	%
	\bibitem{chen2005}
    L. W. Chen, C. M. Ko, and B. A. Li, Phys. Rev. C \textbf{72}, 064309 (2005).
	%
	\bibitem{Abbott2017}
	B. P. Abbott et al. (Virgo and LIGO Scientific Collaborations), Phys. Rev. Lett. \textbf{119}, 161101 (2017).
	%
	\bibitem{Lourenso2020}
	O. Louren\c{c}o et al. Phys. Lett. B \textbf{803}, 135306 (2020).
	\bibitem{Washington2020}
	C. D. Pruitt, R. J. Charity, L. G. Sobotka, M. C. Atkinson, and W. H. Dickhoff, Phys. Rev. Lett. \textbf{125}, 102501 (2020).
	%
	\bibitem{Miller2019}
	M. C. Miller et al., Astrophys. J. Lett. \textbf{887}, L24 (2019). 
	%
	%
	\bibitem{MESA2023}
	D. Becker et al., Eur. Phys. J. A \textbf{54}, 208 (2018).
	%
\end{thebibliography}
\end{document}